\documentclass[preprint,showpacs,preprintnumbers,amsmath,amssymb]{revtex4}

\usepackage{graphicx}
\usepackage{dcolumn}
\usepackage{bm}

\begin{document}



\title{Interlayer correlation between two $^4$He monolayers adsorbed on both sides 
of $\alpha$-graphyne} 
\author{Jeonghwan Ahn}
\author{Sungjin Park}
\author{Hoonkyung Lee}
\author{Yongkyung Kwon} 
\email{ykwon@konkuk.ac.kr}
\affiliation{
Division of Quantum Phases and Devices, School of Physics,
Konkuk University, Seoul 143-701, Korea
}%

\date{\today}


\begin{abstract}
{Path-integral Monte Carlo calculations have been performed to study the $^4$He adsorption 
on both sides of a single $\alpha$-graphyne sheet.
For investigation of the interlayer correlation between the upper and the lower monolayer of $^4$He adatoms, 
the $^4$He-substrate interaction is described by the sum of the $^4$He-C interatomic pair potentials,
for which we use both Lennard-Jones and Yukawa-6 anisotropic potentials.
When the lower $^4$He layer is a C$_{4/3}$ commensurate solid, 
the upper-layer $^4$He atoms are found to form a Kagom\'e lattice structure at a Mott
insulating density of 0.0706~\AA$^{-2}$, and a commensurate solid at an areal density of
0.0941~\AA$^{-2}$ for both substrate potentials. 
The correlation between upper- and lower-layer pseudospins,
which were introduced in Ref.~\cite{kwon13} for two degenerate configurations of three $^4$He atoms in a hexagonal cell,
depends on the substrate potential used;   
With the substrate potential based on the anisotropic Yukawa-6 pair potentials, 
the Ising pseudo-spins of both $^4$He layers 
are found to be anti-parallel to each other 
while the parallel and anti-parallel pseudo-spin alignments
between the two $^4$He layers are nearly degenerate with the Lennard-Jones potentials.
This is attributed to the difference in the interlayer distance,
which is $\sim 4$~\AA~ with the Yukawa-6 substrate potential
but as large as $\sim 4.8$~\AA~with the Lennard-Jones potential. 
}
\end{abstract}


\maketitle

\section{Introduction}
\label{sec:intro}
Among many substrates, graphite has long served as a test bed to investigate  
low-dimensional quantum fluids because of its strong binding of adsorbates. 
Up to seven distinct $^4$He layers were observed on graphite
and each helium layer is considered to be a quasi-two-dimensional quantum system~\cite{zimmerli92}.
The first $^4$He adlayer on graphite shows a commensurate-incommensurate solid
transition as the helium coverage increases~\cite{greywall91,greywall93,crowell96}. 
Recently, a series of theoretical calculations have been
performed to study the $^4$He adsorption on newly-synthesized (or -proposed) low-dimensional 
carbon substrates such as graphene~\cite{gordillo09,kwon12,happacher13}, 
graphynes~\cite{kwon13,ahn14}, carbon nanotubes~\cite{cole00,gordillo08}, 
and fullerene molecules~\cite{kwon10,shin12b,kim13,park14}. 
The phase diagrams of the $^4$He layers adsorbed on graphene were
predicted to be very similar to those of the corresponding layers on graphite;
the monolayer of $^4$He adatoms shows a C$_{1/3}$ commensurate structure at the areal density
of 0.0636~\AA$^{-2}$ and goes through various domain-wall phases before crystallizing into
an incommensurate triangular solid near its completion~\cite{kwon12,happacher13}. 

Graphyne is a two-dimensional (2D) network of $sp$- and $sp^2$-bonded C atoms~\cite{baughman87,coluci04}
which could be permeable to a $^4$He gas unlike graphene.
Despite much experimental effort motivated by some promising theoretical predictions 
for graphyne as new Dirac materials~\cite{malko12,kim12,chen13} 
and high-capacity energy storage materials~\cite{zhang11,hwang12,hwang13},
there has been no successful report yet for fabrication of extended 2D graphynes. 
However, some flakes or building blocks of finite-size graphynes have been synthesized~\cite{tobe03,haley08,diederich10},
leading to a belief that graphynes will be fabricated in the near future. 
On the surface of $\gamma$-graphyne, which is the most stable graphyne structure
according to quantum Monte Carlo calculations~\cite{shin14},
the $^4$He monolayer was predicted to exhibit a richer phase diagram 
than the corresponding layer on graphene or graphite,
including various commensurate and incommensurate structures depending on the helium density~\cite{ahn14}.
Recently one of us performed path-integral Monte Carlo (PIMC) calculations
for $^4$He atoms adsorbed on a AB-stacked bilayer $\alpha$-graphyne~\cite{kwon13},
which is a hybridized honeycomb structure with each hexagon side consisting of one $sp^2$ and two $sp$ C atoms.
It was found that the $^4$He monolayer was in a Mott insulating state at an areal
density of 0.0706~\AA$^{-2}$ while a commensurate solid was realized at 0.0941~\AA$^{-2}$.
Introducing Ising pseudo-spin degrees of freedom for two degenerate configurations
for three $^4$He atoms occupying a hexagonal cell (see Fig. 3 of Ref.~\cite{kwon13}), 
this Mott-insulator to commensurate-solid transition was interpreted 
as a symmetry breaking process from a spin liquid of geometrically-frustrated antiferromagnets to 
a spin-aligned ferromagnet~\cite{kwon13}.

One interesting feature of a 2D carbon structure, such as graphene and graphyne, 
is that it can be suspended in the air~\cite{meyer07} and $^4$He atoms can be coated on both sides.
Noting that some new physics could emerge as a result of interlayer correlation between opposite-side $^4$He layers,
some theoretical studies were recently done for the $^4$He adsorption
on both sides of a single graphene sheet.
Marki\'c {\it et~al.} found that the correlation between two $^4$He clusters
adsorbed on opposite sides of graphene, $5\sim6$~\AA~ apart from each other,
was quite weak as evidenced by peakless pair distribution functions~\cite{markic13}.
A weak correlation between two $^4$He systems on the opposite sides of graphene
was also predicted by Gordillo's diffusion Monte Carlo calculations, 
which showed that the phase diagram of the $^4$He monolayer on graphene
would not be affected by the $^4$He adsorption on the other side~\cite{gordillo14}.  
In this paper we report PIMC study of the $^4$He
adsorption on both sides of a single $\alpha$-graphyne sheet. 
Because $\alpha$-graphyne is more porous than graphene,
$^4$He atoms can penetrate through graphyne to allow physical exchanges among $^4$He atoms on opposite sides.
This could result in stronger interlayer correlation than the corresponding systems on graphene.
We find that $^4$He atoms in a Mott-insulating state 
form a 2D Kagom\'e lattice as a result of the interlayer correlation
when the opposite-side $^4$He layer is a C$_{4/3}$ commensurate solid, 
a ferromagnetic state in a pseudospin terminology.
Effects of the interlayer correlation between two ferromagnetic C$_{4/3}$ solids
are found to depend on the substrate potential used;
the parallel and the antiparallel pseudospin alignments between two $^4$He layers
are nearly degenerate with the substrate potential based on the Lennard-Jones (LJ) $^4$He-C pair potentials
while the antiparallel alignment is favored with the one described by the Yukawa-6 pair potentials.
The vacancy formation in a $^4$He layer on $\alpha$-graphyne is also found to be affected 
by the presence of the opposite-side $^4$He layer. 

In the following section, we outline our approach and some computational details.
The PIMC results along with the related discussions are presented in detail in Sec.~\ref{sec:results}. 
We summarize our findings in Sec.~\ref{sec:conclude}.

\section{Methodology}
\label{sec:method}
In this study, a single $\alpha$-graphyne sheet is set to be at $z=0$. The 
$^4$He-graphyne interaction is assumed to be a pairwise sum of interatomic potentials between
the carbon atoms and a $^4$He atom, 
which has been widely used to describe the interaction between a $^4$He atom and a carbon substrate
~\cite{kwon12,markic13,kwon13,gordillo14,ahn14}. 
For the $^4$He-C interatomic pair potential, 
we employ two anisotropic potentials proposed by Carlos and Cole~\cite{carlos79,carlos80},
{\it i.e.}, a 6-12 LJ potential and a Yukawa-6 potential.
For the computational convenience 
our previous study for the $^4$He monolayer on bilayer $\alpha$-graphyne
was done with only isotropic parts of the LJ pair potential.
However, the original interatomic pair potentials of Carlos and Cole
include anisotropic parts to fit helium scattering data from graphite surfaces.
Even though the inclusion of the anisotropic parts of the interatomic potentials
has little effect on quantum phases displayed by the $^4$He layer
on one side of $\alpha$-graphyne,
it allows some $^4$He atoms to be closer to the substrate,
resulting in stronger correlation between two $^4$He layers on the opposite sides
(the minima of the substrate potential
made of the anisotropic pair potentials are deeper and closer to graphyne than
the corresponding ones based on only isotropic parts of the pair potentials).
This leads to our decision of using the substrate potentials based on fully-anisotropic interatomic pair potentials,
which should give a better description of the interlayer correlation.  
Furthermore, since the LJ and the Yukawa-6 potentials used in this study were based on 
an interaction between helium and $sp^2$-bonded carbon atoms in graphite, 
we tested the sensitivity of our modelling of $^4$He-graphyne potentials to the well depth of the pair potentials.
Although decrease in the well depth yields more fluctuations in $^4$He density distributions,
the density modulations are found to change only little 
and our main results presented below are still, at least qualitatively, valid.
For the $^4$He-$^4$He interaction, 
we use a well-known Aziz potential~\cite{aziz92}.
 
In the discrete path-integral representation, the thermal density matrix at a low temperature 
is expressed by a convolution of $M$ high-temperature density matrices with an imaginary 
time step of $\tau=1/(Mk_BT)$~\cite{ceperley95}. 
While the isotropic parts of $^4$He-C pair potentials along with the $^4$He-$^4$He potential
pair potentials are used to compute the exact two-body density matrices~\cite{ceperley95,zillich05} 
at the high temperature $MT$, their anisotropic parts are treated 
with the primitive approximation~\cite{ceperley95}. 
This is found to give accurate description of both $^4$He-$^4$He and $^4$He-graphyne interactions 
with a time step of $(\tau k_B)^{-1}=80$~K. 
We employ the multilevel Metropolis algorithm to sample the imaginary time paths along with 
permutations among $^4$He atoms as described in Ref.~\cite{ceperley95}. To minimize finite size 
effects, periodic boundary conditions are applied along the lateral directions. 

\section{Results}
\label{sec:results}
The PIMC calculations
were done with a fixed $3 \times 2$ simulation cell with dimensions of $21.01 \times 24.26$~\AA$^2$, 
the same as in our previous study for $^4$He on bilayer $\alpha$-graphyne~\cite{kwon13}.
We focus on the interlayer correlation between two $^4$He layers on the opposite sides of graphyne,
which are either in a Mott-insulating state or a pseudospin-aligned commensurate solid state.
The results obtained with two different substrate potentials, the LJ potential and the Yukawa-6 one, 
are presented separately below.

\subsection{Lennard-Jones substrate potential}
\label{subsec:lj}

\begin{figure}
\includegraphics[width=3.0in]{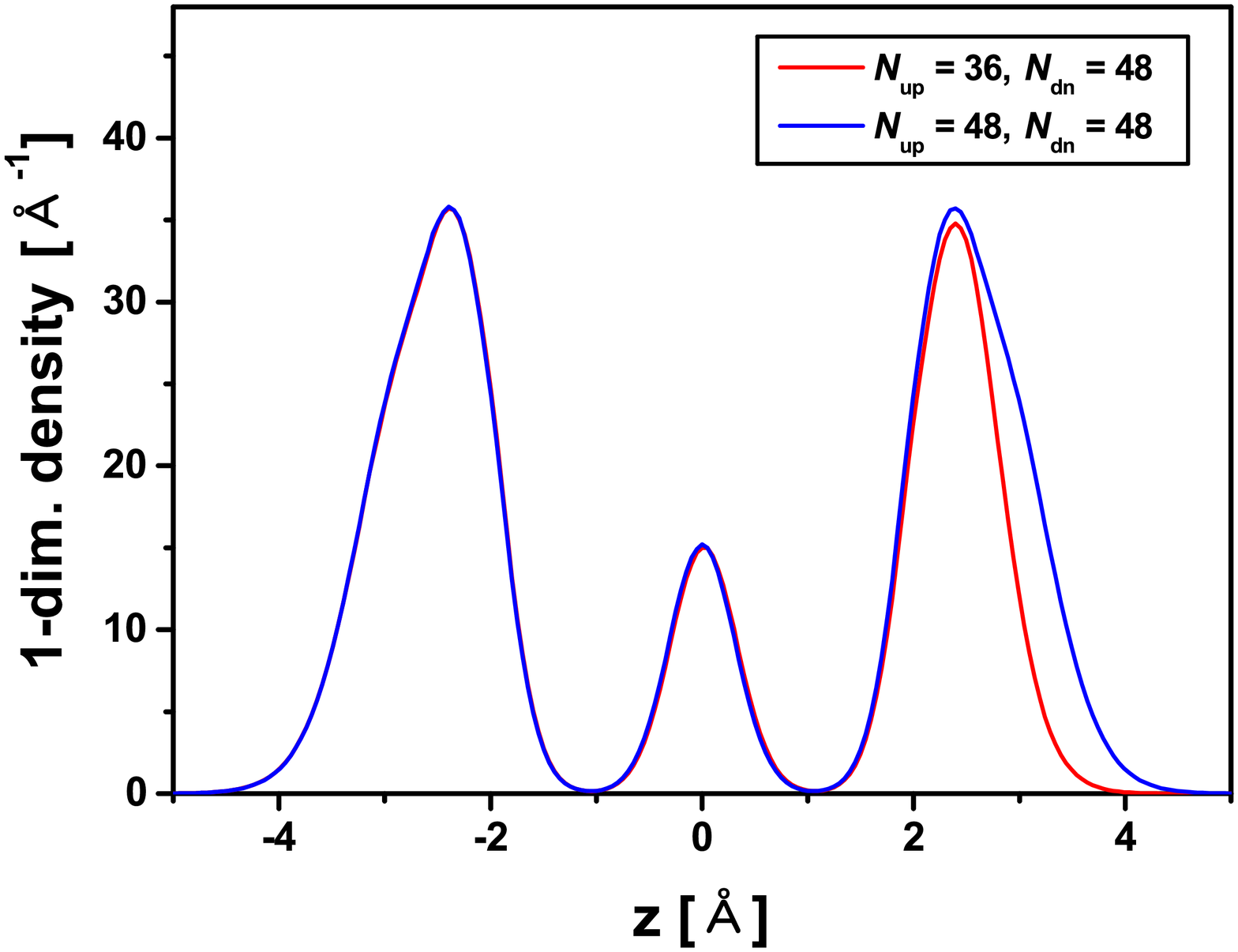}
\caption{(Color online) One-dimensional density distributions of $^4$He atoms 
 adsorbed on both sides of $\alpha$-graphyne
 as a function of the vertical coordinate $z$ perpendicular to the graphyne surface, 
 which were computed with the LJ substrate potential.
 Here $N_{\rm up}$ and $N_{\rm dn}$ represent the number of $^4$He atoms per $3 \times 2$ simulation cell
 in the upper and the lower $^4$He layer, respectively.
 Additional 12 $^4$He atoms per simulation cell are involved to form the zeroth layer around $z=0$ 
 where one $^4$He atom is embedded at every hexagon center (see Ref.~\cite{kwon13}). 
}
\label{fig:1Dden_LJ}
\end{figure}

For PIMC calculations with the LJ substrate potential, we first prepare the $\alpha$-graphyne surface
whose in-plane hexagon center is occupied by a single $^4$He atom and 
whose lower side is coated with a monolayer of $^4$He atoms constituting a C$_{4/3}$ commensurate solid
while each of the in-plane centers is occupied by a single $^4$He atom.
The simulations for the $^4$He adsorptions on the upper side of the prepared graphyne surface 
begin from an initial configuration of $^4$He atoms being randomly distributed
at the distances far away from graphyne.
Figure~\ref{fig:1Dden_LJ} presents one-dimensional (1D) density distributions of $^4$He atoms,
as a function of the vertical coordinate $z$ along the direction perpendicular to the graphyne surface,
for two different combinations of particle numbers per simulation cell. 
Two distinct density peaks, which correspond to the upper and the lower $^4$He layers, 
are observed on the opposite sides of graphyne. 
Note that 36 and 48 $^4$He atoms per simulation cell correspond to the Mott-insulating density
of  0.0706~\AA$^{-2}$ and the C$_{4/3}$ commensurate density of 0.0941~\AA$^{-2}$, respectively. 
The additional density peak at $z=0$ corresponds to the zeroth layer
consisting of $^4$He atoms embedded onto the in-plane hexagon centers,
which was also observed in our previous study for $^4$He adatoms on a bilayer $\alpha$-graphyne~\cite{kwon13}.
Since the peak-to-peak distance between the upper and the lower layers
is estimated to be about 4.8~\AA, one can expect that the van der Waals interaction between $^4$He atoms
on the opposite sides is weakly attractive (note that the Aziz potential we used for the $^4$He-$^4$He interaction
has a minimum value at $r \sim 3.0$~\AA).
In addition, the clear separation between the adjacent density peaks in Figure~\ref{fig:1Dden_LJ}
suggests that exchange couplings among $^4$He atoms in different layers are nearly absent
and any correlation between the upper and the lower layers, if it exists, should stem mostly from
the weakly-attractive $^4$He-$^4$He interaction rather than particle exchanges.    

\begin{figure}
\includegraphics[width=3.5in]{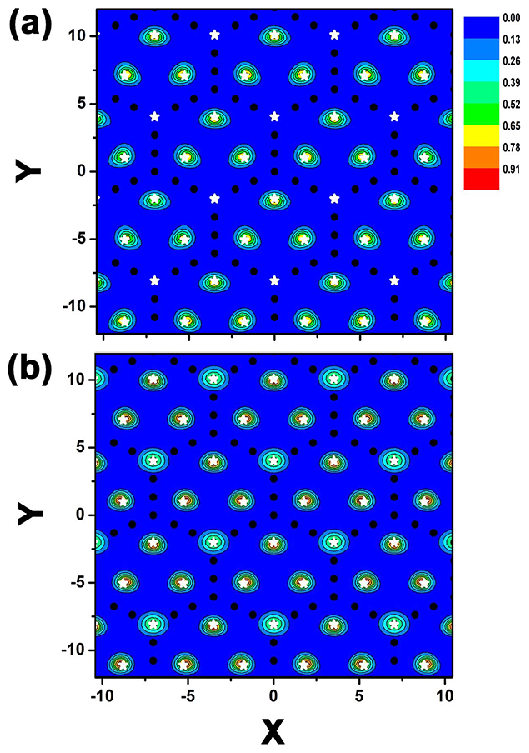}
\caption{(Color online) Contour plots of two-dimensional density distributions of $^4$He atoms
adsorbed on the upper side of $\alpha$-graphyne for upper-layer areal densities 
of (a) 0.0706~\AA$^{-2}$ and (b) 0.0941~\AA$^{-2}$ (red: high, blue: low).
The black dots correspond to the carbon atoms and 
the white stars represent the peak
positions of the lower-layer $^4$He density distribution, which form a C$_{4/3}$ commensurate solid. 
The computations were done at $T=0.5$~K with the LJ substrate potential and the length unit is \AA. 
}
\label{fig:2Dden_LJ}
\end{figure}

Figure~\ref{fig:2Dden_LJ} shows two-dimensional (2D) density distributions of the upper-layer $^4$He atoms,
while the lower-layer density peaks represented by the white stars 
constitute a C$_{4/3}$ commensurate structure with all pseudospins 
being in the spin-up state (see Fig. 3 of Ref.~\cite{kwon13}). 
Here a distinct density peak in each plot represents an occupancy of a single $^4$He atom. 
At an areal density of 0.0706~\AA$^{-2}$, every hexagonal cell of graphyne is seen in Fig.~\ref{fig:2Dden_LJ}(a)
to accommodate three upper-layer $^4$He atoms, which is a manifestation of a Mott-insulating state. 
Without the lower $^4$He layer in a pseudospin-aligned commensurate solid state,
this Mott-insulating state is a nonmagnetic spin liquid of frustrated antiferromagnets
in terms of pseudospin degrees of freedom (see Fig. 2(d) in Ref.~\cite{kwon13}).
However, in the presence of the pseudospin-aligned lower $^4$He layer,
the upper-layer pseudospins are shown in Fig.~\ref{fig:2Dden_LJ}(a) to be 
aligned in the same direction as the lower-layer ones.
Our PIMC simulations at $T=0.5$~K have also produced the antiparallel pseudospin alignment 
between the two $^4$He layers.
This is understood by the fact that  
the parallel alignment is energetically favored only by $\sim 0.3$~K per an upper-layer
helium atom over the antiparallel alignment,
{\it i.e.}, two pseudospin alignments are nearly degenerate.
We here note that the upper-layer $^4$He atoms in a pseudospin-aligned
Mott insulating state of Fig.~\ref{fig:2Dden_LJ}(a)  constitute a 2D Kagom\'e lattice. 
This is also true when the upper-layer pseudospins are aligned in the opposite
direction to the lower-layer ones.
Therefore one can conclude that as a result of the interlayer correlation,
the upper-layer $^4$He atoms form a Kagom\'e lattice structure 
at the Mott insulating density of $0.0706$~\AA$^{-2}$
when the lower $^4$He atoms constitute a pseudospin-aligned C$_{4/3}$ commensurate solid.

The interlayer correlation between two ferromagnetic C$_{4/3}$ commensurate solids
is also analyzed. Figure~\ref{fig:2Dden_LJ}(b) presents the 2D density distribution of the upper-layer
$^4$He atoms at the areal density of $0.0941$~\AA$^{-2}$, where they 
constitute a 4/3 commensurate solid.
The upper-layer pseudospins are seen to be aligned in the same direction as the lower-layer ones.
Similarly to the case of the Mott-insulating state,
the antiparallel pseudospin alignment was also observed in our simulations.
These two pseudospin alignments are more degenerate 
(the parallel alignment was found to be preferred by $\sim 0.11$~K per upper-layer $^4$He atom), 
than the parallel and antiparallel pseudospin alignments between a Mott-insulator and a C$_{4/3}$ solid 
in Fig.~\ref{fig:2Dden_LJ}(a). 
This can be understood by the fact that
unlike $^4$He atoms inside a hexagonal cell, 
upper-layer $^4$He atoms at the vertices of the graphyne hexagons in a C$_{4/3}$ solid state
prefer the other sublattice sites over the ones occupied by the corresponding lower-layer helium atoms. 
We note that once a pseudospin alignment between the two 4/3 commensurate $^4$He solids is established, 
either parallel or antiparallel to each other, 
energy barrier is too large to reverse one alignment to the other. 

\begin{figure}
\includegraphics[width=3.5in]{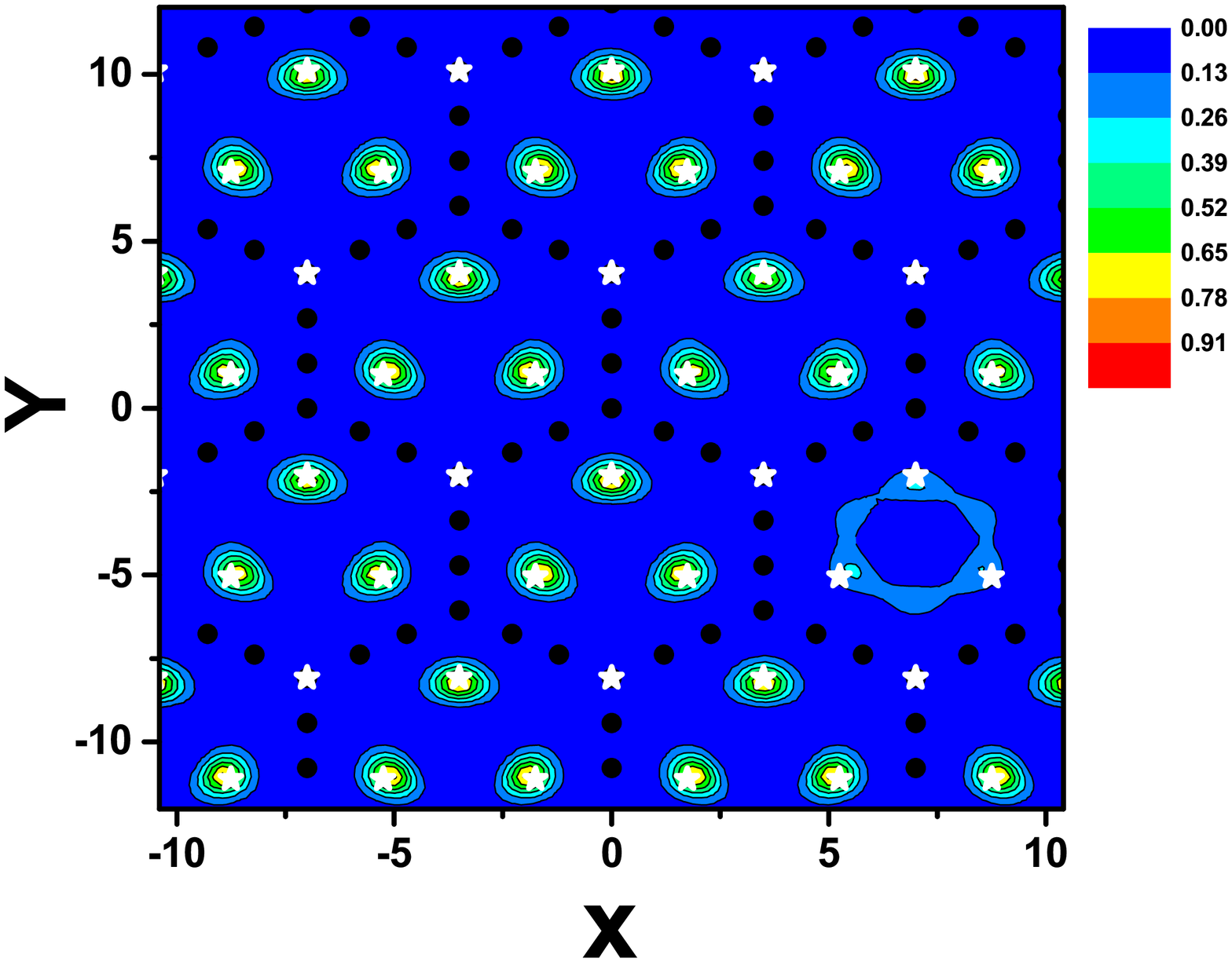}
\caption{(Color online) Contour plot of two-dimensional density distribution of the upper $^4$He layer 
at the areal density of  0.0687~\AA$^{-2}$, which corresponds to one less $^4$He atoms
per our $3 \times 2$ rectangular simulation cell than the Mott-insulating density (red: high, blue: low).
The black dots correspond to the carbon atoms and 
the white stars represent the density peaks
of the lower-layer C$_{4/3}$ commensurate solid. 
The computations were done at $T=0.5$~K 
with the LJ substrate potential and the length unit is \AA.  
}
\label{fig:vac_LJ}
\end{figure}

Now we analyze the effects of the interlayer correlation on the formation of vacancies
especially in a Mott insulator with the Kagom\'e lattice structure. 
Figure~\ref{fig:vac_LJ} shows the 2D density distribution of 35 upper-layer $^4$He atoms, 
one less than the Mott-insulating case, per $3 \times 2$ simulation cell.
So the upper-layer Mott insulator contains one vacancy per simulation cell  
while the lower $^4$He layer is the same 4/3 commensurate solid as in Fig.~\ref{fig:2Dden_LJ}(a).
One can see that
every hexagonal cell, except one, is seen to accommodate three upper-layer $^4$He atoms
and its pseudospin is aligned in the same direction as those 
of the ferromagnetic lower layer.
As shown in Fig.~\ref{fig:vac_LJ},
one cell involving only two upper-layer atoms 
does not show the clear pseudospin alignment.  
This tells us that the Kagom\'e lattice structure is sustained even with the creation of vacancies
but those isolated vacancies are restricted at one triangle of this trihexagonal tiling structure
without hopping to the neighboring sites because of high potential barrier provided by the graphyne surface.

\subsection{Yukawa-6 substrate potential}
\label{subsec:yu6}

\begin{figure}
\includegraphics[width=3.0in]{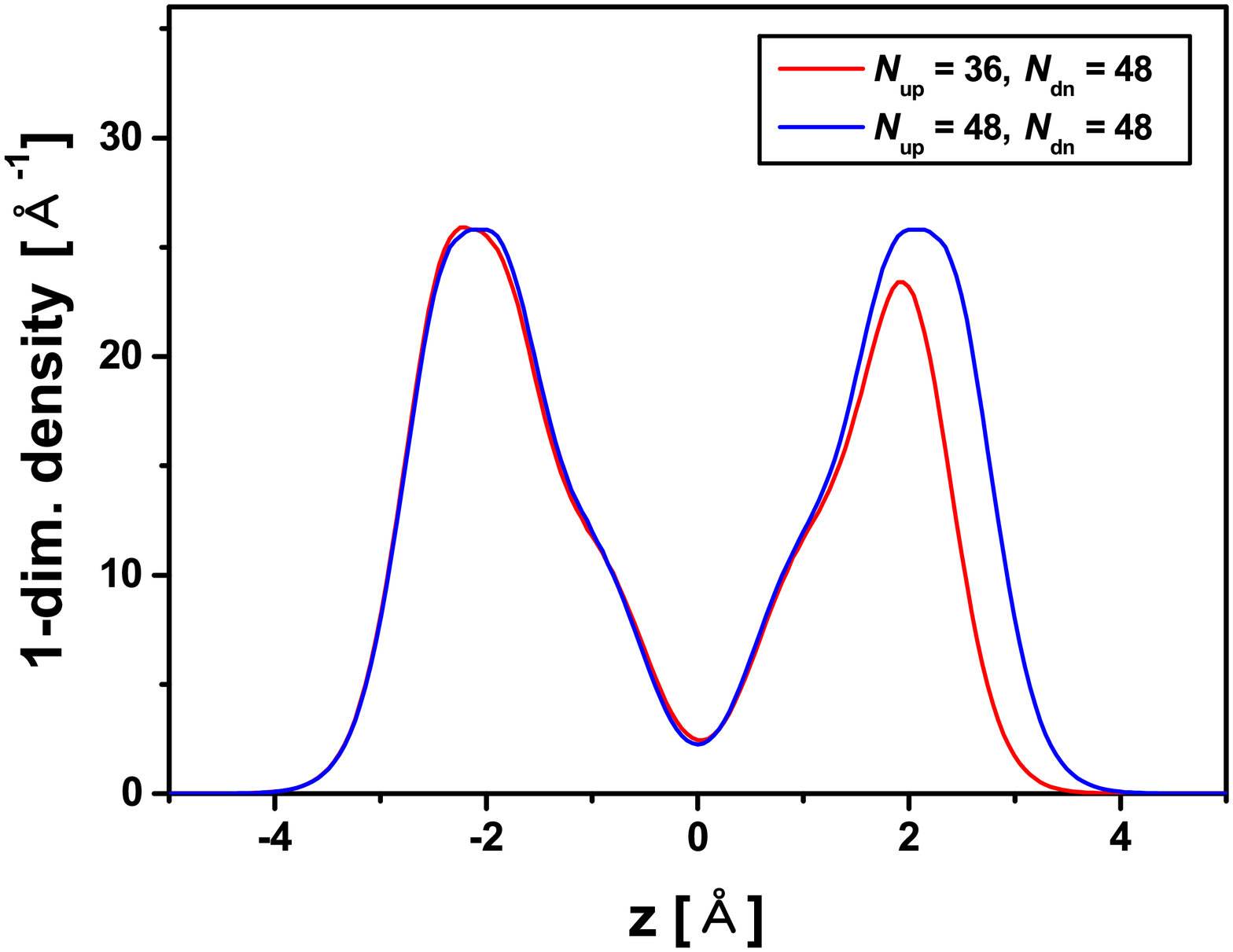}
\caption{(Color online) One-dimensional density distributions of $^4$He atoms 
 adsorbed on both sides of $\alpha$-graphyne
 as a function of the vertical coordinate $z$ perpendicular to the graphyne surface, 
 which were computed with the Yukawa-6 substrate potential.
 Here $N_{\rm up}$ and $N_{\rm dn}$ represent the number of $^4$He atoms per $3 \times 2$ simulation cell
 in the upper and the lower $^4$He layer, respectively. 
}
\label{fig:1Dden_Yu}
\end{figure}

Our PIMC simulations with the Yukawa-6 substrate potential 
start from an initial configuration of $N_{\rm up}$ and $N_{\rm dn}$ $^4$He atoms being distributed randomly 
on the upper and the lower side of $\alpha$-graphyne, respectively.  
Figure~\ref{fig:1Dden_Yu} presents the 1D $^4$He density distributions 
as a function of the vertical coordinate $z$
for two different values of $N_{\rm up}$ while $N_{\rm dn}$ is fixed to 48 per $3 \times 2$ simulation cell.
Unlike Fig.~\ref{fig:1Dden_LJ} for the LJ substrate potential, 
only two density peaks on the opposite sides of graphyne are observed
without the zeroth layer consisting of $^4$He atoms embedded onto the in-plane hexagon centers.
The absence of the zeroth $^4$He layer is attributed to the fact that the Yukawa-6 substrate potential
is more slowly varying near the potential minima, that is, the hexagon centers, 
than the LJ substrate potential.
Note that the Yukawa-6 $^4$He-C pair potential is less repulsive at short distances 
than the LJ interatomic pair potential~\cite{carlos80}.  
The distance between two density peaks is $\sim 4.0$~\AA, 
shorter than the corresponding distance for the LJ substrate potential.
Furthermore, there is significant overlap between the two density peaks
that are broader than those in Fig.~\ref{fig:1Dden_LJ}. 
This indicates large quantum fluctuations of $^4$He adatoms along the vertical direction,
which could result in frequent particle exchanges between these two layers.   

\begin{figure}
\includegraphics[width=3.5in]{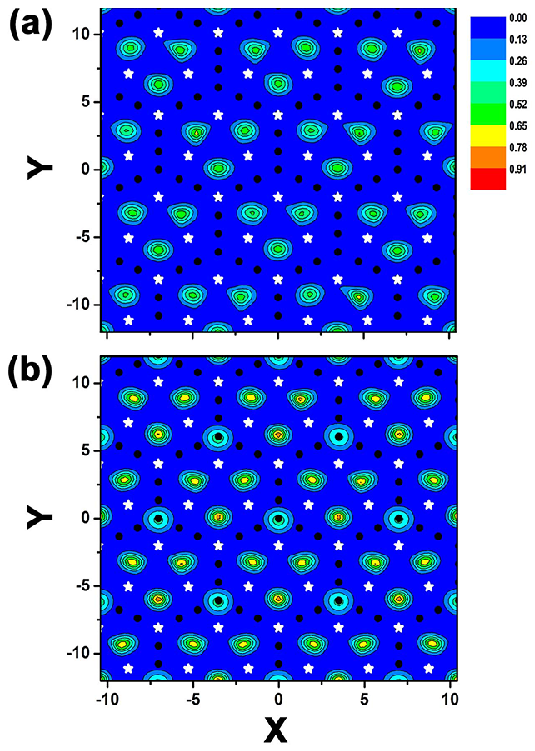}
\caption{(Color online) (a) Contour plots of two-dimensional density distributions of $^4$He atoms
adsorbed on the upper side of $\alpha$-graphyne for upper-layer areal densities 
of (a) 0.0706~\AA$^{-2}$ and (b) 0.0941~\AA$^{-2}$ (red: high, blue: low).
The black dots correspond to the carbon atoms and 
the white stars represent the peak
positions of the $^4$He density distribution of the lower layer whose density is 0.0941~\AA$^{-2}$ for both cases.
The computations were done at $T=0.5$~K 
with the Yukawa-6 substrate potential and the length unit is \AA. 
}
\label{fig:2Dden_Yu}
\end{figure}

Figure~\ref{fig:2Dden_Yu} shows 2D density distributions of the upper-layer $^4$He atoms,
while the lower-layer density peaks are represented by the white stars.
Even with the Yukawa-6 substrate potential, the lower $^4$He layer is seen to
constitute a pseudospin-aligned a C$_{4/3}$ commensurate solid at the areal density of 0.0941~\AA$^{-2}$. 
This provides another confirmation to the conclusion of Ref.~\cite{kwon13}
that most of quantum phases manifested in the $^4$He monolayer on the $\alpha$-graphyne surface, 
such as a Mott insulator, commensurate solids, and pseudospin degrees of freedom, 
are not sensitive to the specifics of the substrate potential but are determined mostly by the surface geometry.
It is also shown in Fig.~\ref{fig:2Dden_Yu}(a) that the upper-layer $^4$He atoms 
under the Yukawa-6 substrate potential are in a Mott-insulating state 
at the areal density of $0.0706$~\AA$^{-2}$ with each hexagonal cell accommodating three $^4$He atoms.
Unlike Fig.~\ref{fig:2Dden_LJ}(a), however, all upper-layer pseudospins are 
aligned in the opposite direction to those of the ferromagnetic lower-layer commensurate solid.
We understand that
the increase in the effective hard-core radii of $^4$He adatoms due to larger quantum fluctuations,
along with a shorter interlayer distance,
causes the antiparallel pseudospin alignment to be favored under the Yukawa-6 substrate potential.
As observed with the LJ substrate potential, 
the upper-layer $^4$He adatoms in the pseudospin-aligned Mott insulating state
constitute a 2D Kagom\'e lattice structure.
The interlayer correlation that favors the antiparallel pseudospin alignment
is more evident between two ferromagnetic C$_{4/3}$ solids.
We observe in Fig.~\ref{fig:2Dden_Yu}(b) that the pseudospins of a upper-layer
4/3 commensurate solid are in spin-down state while the lower-layer pseudospins are in spin-up state.
This antiparallel pseudospin alignment between the two $^4$He adlayers
corresponds to the AB stacking of two triangular solids.
So we conclude that with the Yukawa-6 substrate potential, the AB stacking is preferred to
the AA stacking between two C$_{4/3}$ triangular $^4$He solids
while these two stacking orders are nearly degenerate with the LJ potential as discussed in Sec.~\ref{subsec:lj}.

\begin{figure}
\includegraphics[width=3.5in]{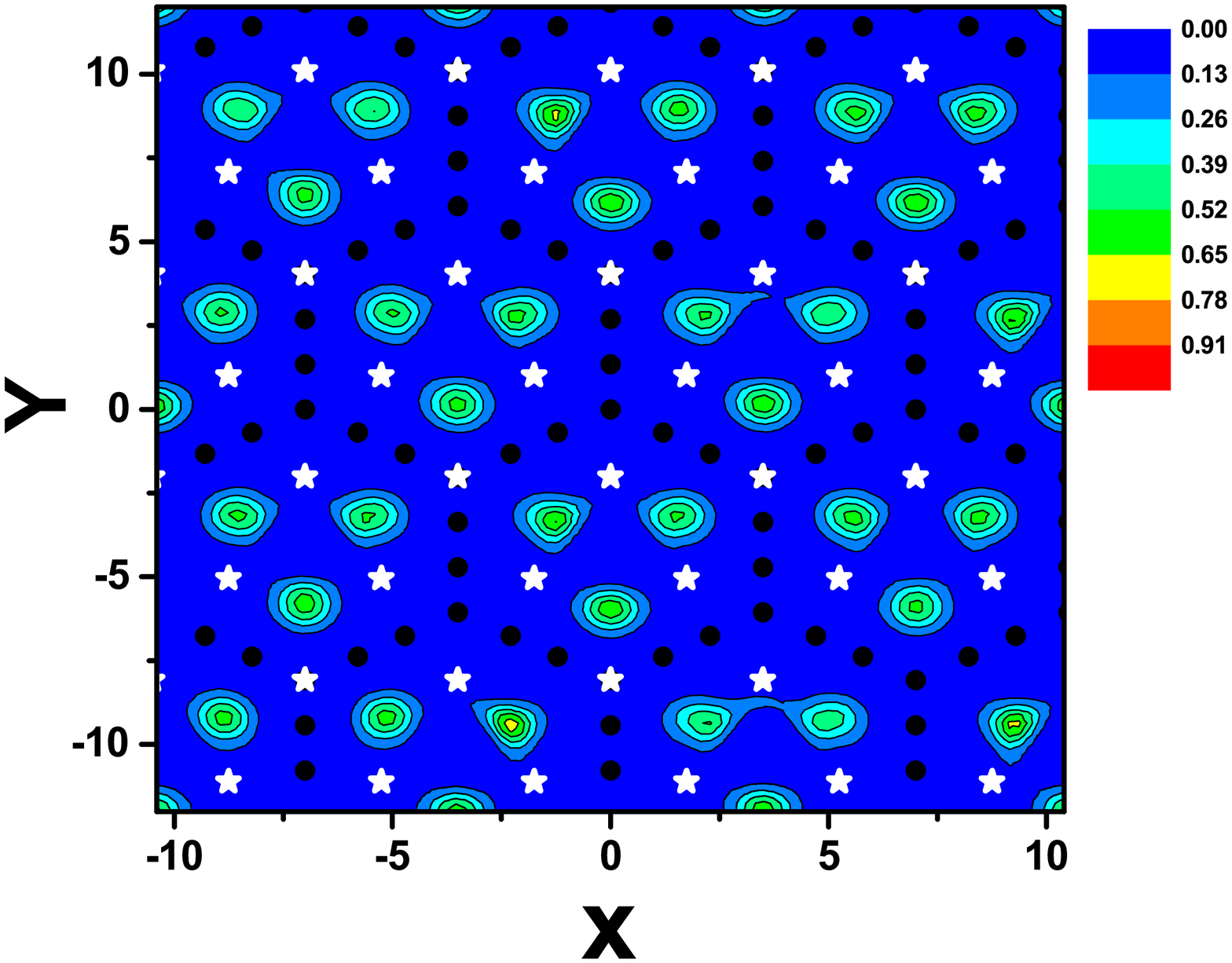}
\caption{(Color online)  Contour plot of two-dimensional density distribution of the upper $^4$He layer 
at the areal density of  0.0687~\AA$^{-2}$, which corresponds to one less $^4$He atoms
per our $3 \times 2$ rectangular simulation cell than the Mott-insulating density (red: high, blue: low).
The black dots correspond to the carbon atoms and 
the white stars represent the density peaks
of the lower-layer C$_{4/3}$ commensurate solid. 
The computations were done at $T=0.5$~K 
with the Yukawa-6 substrate potential and the length unit is \AA. 
}
\label{fig:vac_Yu}
\end{figure}

We now try to create a single vacancy in the upper-layer Mott insulator, {\it i.e.}, the Kagom\'e lattice structure,
by putting only 35 $^4$He atoms on the upper side in our simulation. 
Figure~\ref{fig:vac_Yu} presents the 2D density distribution of the upper-layer $^4$He atoms
along with the lower-layer density peaks of a pseudospin-up C$_{4/3}$ commensurate solid. 
It is shown that there are still three upper-layer density peaks per every hexagonal cell,
corresponding to a configuration of a Mott insulator with all pseudospins being in the down state.  
However, one lower-layer density peak or one white star is missing at
a vertex of the graphyne honeycomb structure (see the bottom right corner), 
which prevents the lower-layer $^4$He atoms from forming a perfect C$_{4/3}$ triangular lattice.
This suggests that when a single vacancy is created in a upper-layer Kagom\'e lattice,
one lower-layer $^4$He atom moves to the upper layer to form a perfect upper-layer lattice structure
while a localized vacancy is created in the lower-layer 4/3 commensurate solid.
This lower-layer vacancy is found at a vertex site on top of a carbon atom
because it is a less favorable site for a $^4$He adatom than a site inside a hexagonal cell.
The layer-to-layer hopping of a vacancy can be understood by a short interlayer distance and
large quantum fluctuations along the vertical direction under the Yukawa-6 substrate potential.

\section{Conclusion}
\label{sec:conclude}
According to our PIMC calculations of using two different $^4$He-substrate potentials, 
$^4$He atoms form distinct layers on both sides of a single $\alpha$-graphyne sheet.
Regardless of the substrate potential used, the upper-layer $^4$He atoms
form a 2D Kagom\'e structure at the Mott-insulating density of 0.0706~\AA$^{-2}$
as a result of the interlayer correlation when the lower layer is a pseudospin-aligned C$_{4/3}$ commensurate solid.
Since the interaction of $^3$He atoms with a substrate or between themselves 
is similar to the corresponding interaction for $^4$He, 
the same Kagom\'e lattice structure is expected to be formed
in the fermionic counterpart of the upper $^4$He layer , {\it i.e.}, 
a $^3$He upper layer adsorbed on $\alpha$-graphyne, when
its lower side is coated with the C$_{4/3}$ commensurate helium solid.
We speculate that some novel phenomena related with a geometrically-frustrated antiferromagnetism
such as quantum spin liquids~\cite{balents10}, could emerge 
in this $^3$He Kagom\'e lattice. 

The interlayer correlation results in different stacking orders between two C$_{4/3}$ 
commensurate triangular solids on the opposite sides of graphyne, depending on the substrate potential;
with the Yukawa-6 potential, the AB stacking (an antiparallel pseudospin alignment
between two $^4$He solids) is found to be favored but both AA (a parallel pseudospin alignment)
and AB stacking configurations are nearly degenerate with the LJ substrate potential.
This is attributed to the difference between two substrate potentials in the interlayer distance 
as well as in the magnitude of quantum fluctuations along the vertical direction.
A more accurate $^4$He-graphyne potential would be required to draw a definite conclusion about
the preferred stacking order of two commensurate triangular $^4$He solids on the opposite sides of $\alpha$-graphyne.

Recent theoretical studies done by Marki\'c {\it et~al.}~\cite{markic13} and Gordillo~\cite{gordillo14}  
reported that the interlayer correlation between two $^4$He systems
adsorbed on both sides of graphene was very weak and quantum phase diagram
of one $^4$He layer would not be affected by the presence of the opposite-side $^4$He layer.
On the other hand, our PIMC calculations have revealed some significant effects 
of the interlayer correlation on the structural properties of the $^4$He monolayers on $\alpha$-graphyne.
This difference is understood to be due to much more porous nature of $\alpha$-graphyne 
than graphene.

\begin{acknowledgments}
This work was supported by Konkuk University.
We also acknowledge the support from the Supercomputing Center/Korea Institute 
of Science and Technology Information with supercomputing resources including technical support 
(KSC-2013-C3-033).
\end{acknowledgments}


\newpage

\end{document}